# Improved Achievable Rates for Regularized Tomlinson-Harashima Precoding in Multiuser MIMO Downlink

Bing Hui
*Student Member, IEEE*
The Graduate School of IT & T
Inha University
Incheon, Korea
huibing_zxo@163.com

Manar Mohaisen
*Student Member, IEEE*
The Graduate School of IT & T
Inha University
Incheon, Korea
lemanar@hotmail.com

KyungHi Chang
*Senior Member, IEEE*
The Graduate School of IT & T
Inha University
Incheon, Korea
khchang@inha.ac.kr

*Abstract*—**Tomlinson-Harashima precoding (THP) is considered as a prominent precoding scheme due to its capability to efficiently cancel out the known interference at the transmitter side. Therefore, the information rates achieved by THP are superior to those achieved by conventional linear precoding schemes. In this paper, a new lower bound on the achievable information rate for the regularized THP scheme under additive white Gaussian noise (AWGN) channel with multiuser interference is derived. Analytical results show that the lower bound derived in this paper is tighter than the original lower bound particularly for a low SNR range, while all lower bounds converge to $0.5 \cdot \log_2(6 SNR/\pi e)$ as $SNR \to \infty$.**

## I. INTRODUCTION

Multiple-input multiple-output (MIMO) communication techniques have been an important research topic due to their potential for high capacity, increased diversity, and interference suppression. For applications such as wireless LANs and cellular telephony, MIMO systems will likely be deployed in environments where a base station (BS) simultaneously communicates with many users. As a result, the study of multiuser MIMO (MU-MIMO) systems has recently emerged as an important research topic. Such systems have the potential to exploit the high capacity achievable by MIMO processing and combine this with the benefits of space division multiple access (SDMA).

In MU-MIMO downlink, spatial multiplexing schemes are adopted at the transmitter, where parallel transmission of independent data streams introduces severe interlayer interference. Since the mobiles are decentralized and non-cooperative, conventional MIMO detection algorithms, such as zero-forcing (ZF), minimum mean square error (MMSE) or decision-feedback detection (DFE), cannot work efficiently. An alternative solution is dirty paper coding (DPC), where the detection structure is moved from the receiver side to the transmitter side. Thus, as the streams of different users are perfectly known at the transmitter, i.e., BS, they can be cancelled out such that each user receives only his own stream.

DPC was first described by Costa for the Gaussian interference channel [1], where the capacity of an interference channel (and the interfering signals are known at the transmitter) is shown to be the same as that of the interference-free channel. Recently, DPC has emerged as a building block in multiuser broadcast over MIMO channels, as initiated in [2].

The simplest precoding algorithm for MU-MIMO systems is the well-known matrix inversion, i.e., zero-forcing precoding. By treating the transmitted signal vector with the pseudo-inverse of the channel matrix, the mutual interference among users is perfectly cancelled out [3]. Nevertheless, for an ill-conditioned channel, the transmit power is not fairly distributed among users, resulting in degradation in the system performance. Regularized channel inversion, i.e., MMSE precoding, avoids the noise amplification by improving the conditionality of the channel matrix. Although the sum capacity of linear MMSE precoding is linear in terms of the number of users, it is still far from the Shannon sum capacity of MIMO systems [4].

Tomlinson-Harashima precoding (THP) achieves close to capacity sum rates [5]. The performance and capacity of THP based on the ZF criterion (THP-ZF) are shown to be superior to those of the linear precoding algorithms. Thus, employing the THP with the MMSE performance-based criterion (THP-MMSE) achieves close to the MIMO capacity in absence of interference. In this paper, we are interested in the achievable sum information rates by the THP-MMSE algorithm due to its potential to increase the channel capacity.

In [6], a lower bound on the achievable information rates by THP-MMSE is given. In this paper, a new lower bound on the achievable information rate by THP-MMSE is derived. Our derived lower bound is shown to be tighter than that given in [6] for a low signal to noise ratio (SNR), while both bounds converge at a high SNR.

This work was supported by the Korea Research Foundation Grant funded by the Korean Government (MOEHRD) (KRF-2007-331-D00296(I0048)).

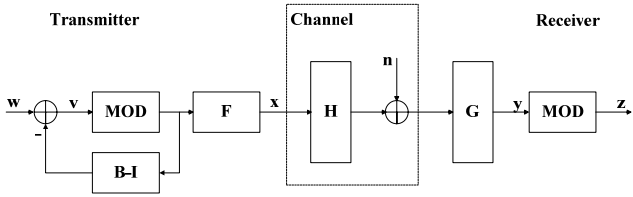

Figure 1. Structure of MU-MIMO system employing Tomlinson-Harashima Precoding

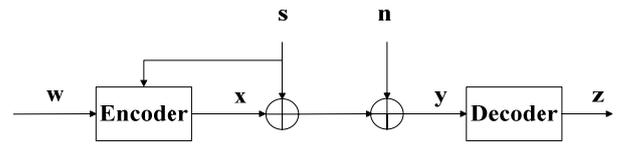

Figure 2. Structure of MU-MIMO system employing dirty paper coding

The remainder of this paper is organized as follows. First, Tomlinson-Harashima precoding is reviewed in Section II. In Section III, we derive the new bound on the achievable information rate by THP-MMSE. Numerical results are given in Section IV. Finally, conclusions are drawn in Section V.

## II. TOMLINSON-HARASHIMA PRECODING FOR MULTIUSER MIMO SYSTEMS

In this paper, we assume that the transmitter has perfect knowledge of the channel state information (CSI) via feedback from users. There are $N$ transmit antennas deployed in the base station (BS), and $M$ non-cooperative mobile stations (MS) simultaneously communicating with the BS. Each MS has one receive antenna. The structure of communication system using THP in multiuser MIMO downlink is shown in Figure 1. The symbol vector $\mathbf{w} = (w_1, w_2, \ldots, w_M)^T$ is the data vector to be transmitted, where $w_i$ corresponds to $MS_i$. We also assume that all the transmitted symbols are independent and they have unit average power, i.e., $E[\mathbf{w}\mathbf{w}^H]=\mathbf{I}$, where $\mathbf{I}$ denotes the identity matrix.

In Figure 1, the THP transmitter consists of a feed-forward filter $\mathbf{F}$ and a feedback filter (**B-I**). The feedback filter (**B-I**) transforms the interference into a causal form, thus, the interference can be eliminated by the feedback filter (**B-I**). Note that the diagonal elements of the lower triangular matrix $\mathbf{B}$ should be all 1s. Both the feed-forward and feedback filters are obtained via the LQ decomposition of the channel matrix $\mathbf{H}$, which results in a unitary matrix $\mathbf{Q}$ and a lower triangular matrix $\mathbf{L}$ such that H=LQ, F=Q$^{-1}$ and GL=B, where $\mathbf{G}$ is a diagonal matrix whose elements are the weighting factors at the receivers.

The function of MOD shown in Figure 1 is the non-linear modulo operator which is described as follows.

$$f_t(y) = y - \left\lfloor \frac{y+t/2}{t} \right\rfloor \times t \qquad (1)$$

This modulo function is a mapping from the real numbers R to [-$t$/2, $t$/2] where $t$ is a positive number.

## III. ACHIEVABLE INFORMATION RATES FOR TOMLINSON-HARASHIMA PRECODING

The MU-MIMO system employing THP can be considered the original DPC with non-linear modulo operations at both the transmitter and the receiver. As shown in Figure 2, the channel input symbol vector $\mathbf{x}$ is distorted by the interference $\mathbf{s}$. Additive white Gaussian noise (AWGN) $\mathbf{n}$ is added at the receiver side to produce the channel output $\mathbf{y}$.

For any $\mathbf{w}$ whose elements are independent and identically distributed (i.i.d.) with a uniform distribution in [-$t$/2, $t$/2], the encoder output is given by

$$x = f_t(w - \alpha s) = [w - \alpha s] \bmod t \qquad (2)$$

where that $x \in [-t/2, t/2]$. If the transmitted vector $\mathbf{w}$ is an i.i.d. sequence with a uniform distribution on [-$t$/2, $t$/2], then $\mathbf{x}$ is also i.i.d. with a uniform distribution on [-$t$/2, $t$/2]. As the alphabet size increases, for i.i.d. pulse amplitude modulation (PAM), the total transmit power $P_T$ converges to the second moment of a uniform random variable over [-$t$/2, $t$/2]. that is,

$$p_t = E[x^H x] = t^2/12 \qquad (3)$$

In this paper, $t$ is chosen based on this large alphabet PAM approximation of $P_T$. Thus, to satisfy the power constraint, $t$ is chosen as

$$t = \sqrt{12 P_T} \qquad (4)$$

The AWGN noise at the receivers is assumed to be mutually independent with variance of $P_N = E[n^2] = \sigma_n^2$. Thus, the SNR in this paper can be defined as

$$SNR = \frac{P_T}{P_N} = \frac{t^2}{12\sigma_n^2} \qquad (5)$$

Finally, the receiver computes the following,

$$z = f_t(\alpha y) = [\alpha y] \bmod t \qquad (6)$$

Then, the resulting channel is a mod-$t$ additive noise channel described by the lemma in [7].

*Definition 1 (Mod-t channel)*: The channel from $\mathbf{w}$ to $\mathbf{z}$ described in Figure 2, (2), and (6) is equivalent to the mod-$t$ channel in distribution

$$z = f_t(w + n') = [w + n'] \bmod t \qquad (7)$$

with

$$n' = f_t(\alpha n) = [\alpha n] \bmod t \qquad (8)$$

The factor $\alpha$, which is related to the MMSE precoding, has been derived by Costa [1] as following,

$$\alpha = \sqrt{\frac{P_T}{P_T + P_N}} = \sqrt{\frac{SNR}{1+SNR}} \quad (9)$$

In the case of THP-MMSE, the interference cannot be fully eliminated, since the MMSE criterion achieves a compromise between noise amplification and interference cancellation. Assume that the remaining nonzero interference **s'** is Gaussian distributed with zero mean, and variance of $\alpha_s^2$. Thus, if we define $n_{MMSE}=s'+n$ as the real noise in the system, then $n_{MMSE}$ is also Gaussian distributed with zero mean, and variance of $\sigma^2 = \sigma_s^2 + \sigma_n^2$.

Based on (7) and (8), the output of the THP-MMSE system is given by

$$z = [w + n'] \bmod t \quad (10)$$

where

$$n' = f_t(\alpha n_{MMSE}) = [\alpha n_{MMSE}] \bmod t \quad (11)$$

The mutual information between the input and output of this channel is given by

$$I(w,z) = h(z) - h(z|w), \\ = h([w+n'] \bmod t) - h([n'] \bmod t | w) \quad (12)$$

where $h(.)$ denotes the differential entropy.

A lower bound on capacity $C_{MMSE-THP}$ can be found by considering the elements of **w**, which are independent and uniformly distributed on $[-t/2, t/2]$. For such a **w**, the output **z** is i.i.d. uniform on $[-t/2, t/2]$. Thus,

$$h(z) = \log_2(t) \quad (13)$$

Removing conditioning always increases the entropy, so

$$h([n'] \bmod t | w) \leq h([n'] \bmod t) \\ \leq h([\alpha n_{MMSE}] \bmod t) \quad (14)$$

Here, the range of $[\alpha n_{MMSE}] \bmod t$ is $[-t/2, t/2]$, and $\alpha n_{MMSE}$ is Gaussian distributed with zero mean and variance of $(\alpha\sigma^2)$. Thus, an upper bound on $[\alpha n_{MMSE}] \bmod t$ is the differential entropy for a truncated Gaussian distribution with the range of $[-t/2, t/2]$, zero mean and variance of $(\alpha\sigma^2)$. Assuming that $d$ has the same distribution as $\alpha n_{MMSE}$, it follows that the probability density function (PDF) of $d$ is

$$f(d;0,(\alpha\sigma)^2,-t/2,t/2) = \frac{\frac{1}{(\alpha\sigma)^2}\varphi\left(\frac{d}{\alpha\sigma}\right)}{\Phi\left(\frac{t}{2\alpha\sigma}\right) - \Phi\left(\frac{-t}{2\alpha\sigma}\right)} \\ = \frac{\frac{1}{(\alpha\sigma)^2}\varphi\left(\frac{d}{\alpha\sigma}\right)}{\text{erf}\left(\frac{t}{2\sqrt{2}\alpha\sigma}\right)} \quad (15)$$

where $\varphi(\cdot)$ is the probability density function of the standard Gaussian distribution, $\Phi(\cdot)$ is the cumulative distribution function, and $\text{erf}(\cdot)$ is the error function. Therefore,

$$h([\alpha n_M] \bmod t) \leq h(d) \quad (16)$$

where

$$h(d) = -\int_{-t/2}^{t/2} f(d) \ln[f(d)] d(d) \\ = \frac{\ln\left[\alpha\sigma \cdot \text{erf}\left(\frac{t}{2\sqrt{2}\alpha\sigma}\right)\right]}{\text{erf}\left(\frac{t}{2\sqrt{2}\alpha\sigma}\right)} \cdot \int_{\frac{-t}{2\alpha\sigma}}^{\frac{t}{2\alpha\sigma}} \varphi\left(\frac{d}{\alpha\sigma}\right) d\left(\frac{d}{\alpha\sigma}\right) \\ - \frac{1}{\text{erf}\left(\frac{t}{2\sqrt{2}\alpha\sigma}\right)} \cdot \int_{\frac{-t}{2\alpha\sigma}}^{\frac{t}{2\alpha\sigma}} \varphi\left(\frac{d}{\alpha\sigma}\right) \cdot \ln\left(\frac{d}{\alpha\sigma}\right) d\left(\frac{d}{\alpha\sigma}\right) \quad (17)$$

We next define $u = d/(\alpha\sigma)$. Then, $u$ is standard Gaussian distributed on $[-t/(2\alpha\sigma), t/(2\alpha\sigma)]$. So, (17) can be reduced further, as shown in (18)

$$h(d) = \ln\left[\alpha\sigma \cdot \text{erf}\left(\frac{t}{2\sqrt{2}\alpha\sigma}\right)\right] \\ - \frac{1}{\text{erf}\left(\frac{t}{2\sqrt{2}\alpha\sigma}\right)} \cdot \int_{\frac{-t}{2\alpha\sigma}}^{\frac{t}{2\alpha\sigma}} \varphi(u) \cdot \ln(u) d(u), \\ \leq \ln\left[\alpha\sigma \cdot \text{erf}\left(\frac{t}{2\sqrt{2}\alpha\sigma}\right)\right] + \ln(\sqrt{2\pi}) \\ + \frac{1}{2\text{erf}\left(\frac{t}{2\sqrt{2}\alpha\sigma}\right)} \cdot \int u^2 \varphi(u) d(u) \\ = \ln\left[\alpha\sigma \cdot \text{erf}\left(\frac{t}{2\sqrt{2}\alpha\sigma}\right)\right] + \ln(\sqrt{2\pi}) + \frac{1}{2\text{erf}\left(\frac{t}{2\sqrt{2}\alpha\sigma}\right)} \quad (18)$$

Note that the unit of (18) is nats. By changing the base of the logarithm, we obtain,

$$h(d) = \log_2\left[\alpha\sigma \cdot \text{erf}\left(\frac{t}{2\sqrt{2}\alpha\sigma}\right)\right] + \log_2(\sqrt{2\pi}) \\ + \frac{1}{2\text{erf}\left(\frac{t}{2\sqrt{2}\alpha\sigma}\right)} \log_2 e \quad (19)$$

By combining (5), (9), (12), (13), (14), (17), (18) and (19), the lower bound for THP-MMSE is give by

$$C_{MMSE-THP} \geq \frac{1}{2}\log_2[12(1+SNR')] - \log_2\left[\text{erf}\left(\sqrt{\frac{3}{2}(1+SNR')}\right)\right] \\ - \log_2(\sqrt{2\pi}) - \frac{1}{2\text{erf}\left(\sqrt{\frac{3}{2}(1+SNR')}\right)} \log_2 e \quad (20)$$

Note that $SNR'$ is the ratio between the transmit power and the real noise. Hence,

$$SNR' = \frac{P_T}{P_{N_M}} = \frac{t^2}{12(\sigma_s^2 + \sigma_n^2)} \qquad (21)$$

## IV. NUMERICAL RESULTS

The well-known AWGN channel capacity is equivalent to the mutual information for an i.i.d. non-truncated Gaussian random variable **w**. Thus, it is given by

$$C = \frac{1}{2}\log_2(1+SNR) \qquad (22)$$

In [6], a lower bound on the achievable information rate for multidimensional lattice quantization is shown, and it is the lower bound for THP-MMSE in one-dimensional case. The formula of this lower bound is

$$Lowerbound = \frac{1}{2}\log_2\left[\frac{6(1+SNR)}{\pi e}\right] \qquad (23)$$

For the purpose of comparison, AWGN channel capacity given by (22), our new lower bound given by (20), and the lower bound given by (23) are plotted in Figure 3. Our new bound is marked as "New Bound", and the lower bound given by (23) is marked as "Original Bound".

From Figure 3, we find that the new bound has better performance than the original bound in [6], while both equations converge to 0.5·log2(6*SNR*/π*e*), as *SNR*→∞. At a high SNR, the achievable capacity suffers from the well-known "shaping loss", which is $10\log_{10}(\pi e) \approx 1.53$ dB.

## V. CONCLUSIONS

In this paper, a new lower bound on the achievable information rate by THP-MMSE algorithm is derived. The introduced equation gives a tighter lower bound relative to the original lower bound, particularly at a low SNR. As a close-to-capacity dirty paper coding scheme, the THP-MMSE algorithm was proven herein to have approximately a 1.53 dB capacity lag due to the "shaping loss" at a high SNR.

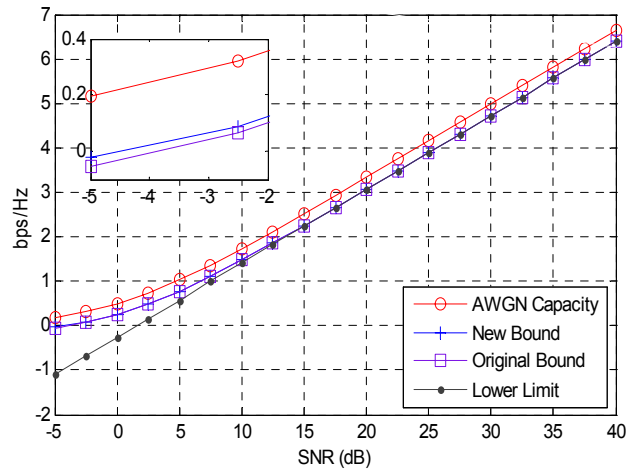

Figure 3. Achievable information rates under the AWGN channel.